\documentclass[FBSedit,FBSmath,ecsub]{FBSsuppl}
\usepackage{amsfonts}
\usepackage{amssymb}
\usepackage{epsfig}
\def\tri{{{}^3{\rm H}}}
\def\het{{{}^3{\rm He}}}

\title{Three-Nucleon Electroweak Capture Reactions}
\author{L.E. Marcucci\instnr{1,2}, 
M. Viviani\instnr{2,1}, 
A. Kievsky\instnr{2,1}, 
S. Rosati\instnr{1,2}, 
R. Schiavilla\instnr{3,4}}
\instlist{Physics Department, University of Pisa, Via Buonarroti 2,
          I-56100 Pisa, Italy
          \and
          INFN, Sezione di Pisa, Via Buonarroti 2, I-56100 Pisa,
          Italia 
          \and
          Jefferson Laboratory, Newport News, Virginia 23606, USA
          \and
          Physics Department, Old Dominion University, Norfolk, 
          VA 23529, USA}
\begin{document}

\maketitle
\begin{abstract}
Recent advances in the study of the $p-d$ radiative and $\mu-\het$ weak
capture processes are presented and discussed. The three-nucleon bound 
and scattering states are obtained using the 
correlated-hyperspherical-harmonics method, with realistic Hamiltonians 
consisting of the Argonne $v_{14}$ or Argonne $v_{18}$ 
two-nucleon and Tucson-Melbourne or Urbana IX three-nucleon 
interactions. The electromagnetic and weak transition operators 
include one- and two-body contributions. The theoretical 
accuracy achieved in these calculations allows for interesting
comparisons with experimental data.
\end{abstract}
\
\section{Introduction}
\label{intro}
A number of electromagnetic (EM) and weak transitions in light nuclei 
have interesting astrophysical implications. They are also very important
for an understanding of nuclear structure and dynamics. 
The theoretical description of these processes
requires the knowledge of the initial and final 
nuclear states, which can be either bound or in the continuum. 
Furthermore, it is important to use a realistic model for the EM and weak
current operators. 
The three-nucleon systems, in particular, provide
a unique ``laboratory" due to the capability, achieved in the last
few years, of obtaining very accurate bound and continuum 
wave functions. The accuracy of the calculated three-nucleon wave
functions has been verified by comparing results for a
variety of observables obtained by a number of groups using
different techniques~\cite{Kie98}.  At present, good overall
agreement exists between the theoretical and experimental $N-d$
elastic and inelastic scattering observables (a notable exception, however,
is the $A_y$ analyzing power at low energies)~\cite{Glo96,KRTV96}.
Therefore the study of EM and weak transitions in the three-nucleon
system does not suffer of uncertainties related to the computation of 
the nuclear wave functions 
and it is a direct test of the nuclear Hamiltonian $H$ from which these
wave functions are obtained, and of the model used to
describe the nuclear currents.  Since the nuclear EM current is
related to $H$ through current conservation, it is clear that
the two topics are inter-related.  
Furthermore, it is interesting to understand whether 
relativistic corrections
as well as $\Delta$-isobar and additional sub-nucleonic degrees of
freedom play a role in these processes. 

The model for the nuclear EM and weak current considered here 
has been recently reviewed in 
Refs.~\cite{SPR89,SR91,CS98} and has been tested in numerous  
few-nucleon processes. It includes
one- and two-body operators. 
The one-body operators are obtained
directly from the non-relativistic limit of the covariant
single-nucleon vector and axial currents. 
In the  study of the muon capture, the contribution coming from the
induced pseudo-scalar term of the nucleon axial current has to be included
(it gives a negligible contribution to $\beta$-decay processes). 
However, the experimental value of the corresponding form factor
$G_{PS}(q_\sigma^2)$ is rather uncertain.  Assuming 
pion-pole dominance, the partially conserved axial current
(PCAC) hypothesis, and the Goldberger-Treiman
relation, $G_{PS}$ is predicted to be~\cite{Mar01,Mar02,Hem95,Wal75,Wal95}
\begin{equation}
  G_{PS}^{PCAC}(q_\sigma^2) = -
  {2\,m_\mu\,m_N \over m_\pi^2 + q_\sigma^2 }\,
  G_{A}(q_\sigma^2) \ ,\label{eq:gps}
\end{equation}
where $q_\sigma$ is the four-momentum transferred to the nuclear system,
$m_N$, $m_\mu$ and $m_\pi$ indicates the nucleon, muon and
pion mass, respectively, and $G_A$ is the axial form factor. 
In our calculation, we have assumed 
\begin{equation}
  G_{PS}(q_\sigma^2) = R_{PS} \; G_{PS}^{PCAC}(q_\sigma^2) 
  \ ,\label{eq:gps2}
\end{equation}
where $R_{PS}$ is a parameter which has been varied
to study the sensibility of our results to this form factor and
to investigate to which extent $G_{PS}^{\rm expt}=G_{PS}^{\rm PCAC}$.
However, most of the calculations have been performed with the choice
$R_{PS}=1$.

The two-body EM current is separated in two terms. There 
is a ``model-independent" part which is constructed consistently
with the nucleon-nucleon interaction, in order to satisfy the
current conservation relation~\cite{R85}. Current conservation 
is only approximate for the momentum dependent terms. 
The second part includes
``model-dependent" contributions which come from the
$\rho\pi\gamma$ and $\omega\pi\gamma$ processes and the $\Delta$
degrees of freedom. The latter contribution is included in the
current and in the nuclear medium in an approximate way, by
following the procedure described in Ref.~\cite{SWPC92}. 
The two-body weak vector current is then obtained from the isovector
part of the EM current, in accordance with the
conserved-vector-current hypothesis.

Two-body terms have been taken into account in both
the axial charge and current operators.  The two-body axial charge
operator has been obtained consistently with the two-nucleon
interaction model, following the methods of Ref.~\cite{Kir92}.
The two-body axial current operators are derived from
a meson-exchange model, including $\pi$- and $\rho$-exchanges
and the $\rho\pi$-transition processes, as well as  
$\Delta$-isobar excitation~\cite{Mar01,Sch98}.
The latter process gives the dominant contribution.  However, its
magnitude depends critically on the value adopted for the $N\Delta$
axial coupling constant $g_A^*$.  In the quark model, $g_A^*$ 
is related in a simple way to the axial coupling constant of the
nucleon $g_A$ ($g_A\approx 1.26$).  However, given the uncertainties inherent
in quark model predictions, a more reliable estimate for $g_A^*$
is obtained by adjusting its value to reproduce the experimental value
of the Gamow-Teller matrix element in tritium
$\beta$-decay~\cite{Mar01,Sch98}.  In this way, the model dependence of
the weak axial current is significantly reduced, as shown by previous
studies of proton weak captures on $^1$H~\cite{Sch98} and $^3$He~\cite{Mar01}.
It is important to note that the value of $g_A^*$ depends on how the 
$\Delta$-isobar degrees of freedom are treated. 
Moreover, given the procedure used to determine $g_A^*$, the latter 
cannot be naively interpreted as the $N\Delta$ axial coupling 
constant, since the contributions of additional resonances not 
included in the present study will contaminate its value. 

The $\het$, $\tri$  bound and the $p-d$ continuum wave functions have
been calculated by expanding on a basis of pair-correlated
hyperspherical harmonic (PHH) functions~\cite{KRV93}. Such a
technique has been proven to be very accurate. Various $n-d$ elastic
scattering observables calculated by solving the Faddeev equations 
~\cite{Glo96} are in remarkable agreement with the
corresponding results obtained with the PHH technique. For example,
the phase shift and mixing angle parameters calculated by the two
methods at the center-of-mass (c.m.) energy $E_{\rm c.m.}=2$ MeV have
been found to 
differ at level of 
$0.1$\%, at most~\cite{Kie98}.  It should be pointed
out that with the PHH technique the inclusion of the
Coulomb potential, clearly very important in the energy
range considered here, does not present any difficulty.

This paper is organized as follows. The $p-d$
radiative capture and the muon capture on $\het$ are discussed in
Sect.~\ref{sec:pd} and~\ref{sec:mu}, respectively.
The last section contains a summary and some concluding remarks.

\section{Radiative $p-d$ Capture}
\label{sec:pd}

\begin{figure}[t]
\centerline{
\epsfig{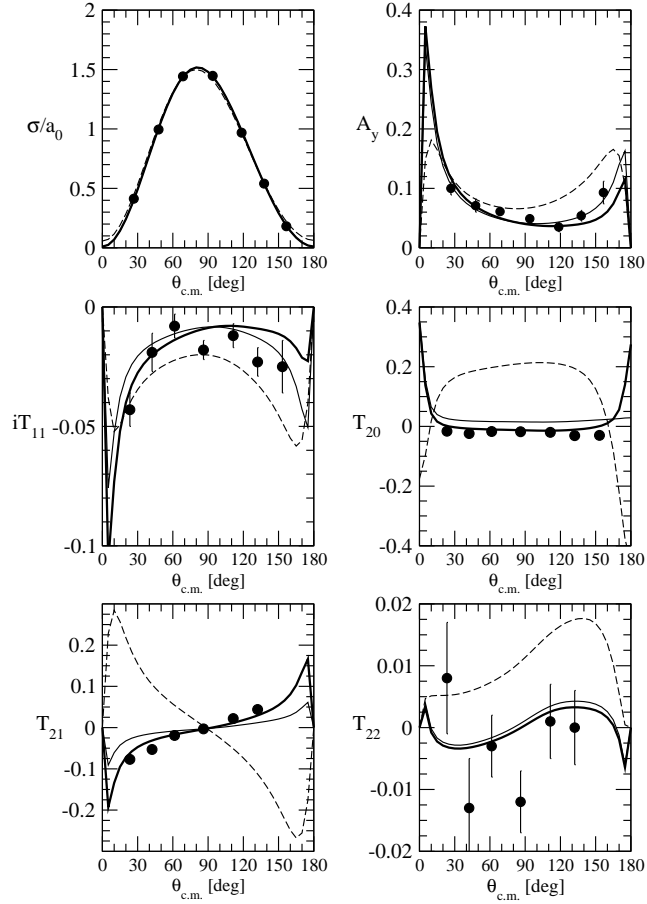}}
\caption{Differential cross section, proton vector analyzing power, and the
four deuteron tensor analyzing powers for 
$p-d$ capture at $E_{\rm c.m.}=3.33$ MeV,
obtained with the AV18/UIX Hamiltonian model 
and one-body only (thick dashed lines)
or both one- and two-body currents (thin solid lines), 
are compared with the
experimental values of Ref.~\protect\cite{Knu92}.  The results
obtained in the LWA approximation  for the spin-flip $E_1$ RME's are
also shown (thick solid lines). In the first panel, $a_0=\int d\Omega\;
\sigma/(4\pi)$.} 
\label{fig:pd3MeV}
\end{figure}

\begin{figure}[t]
\centerline{
\epsfig{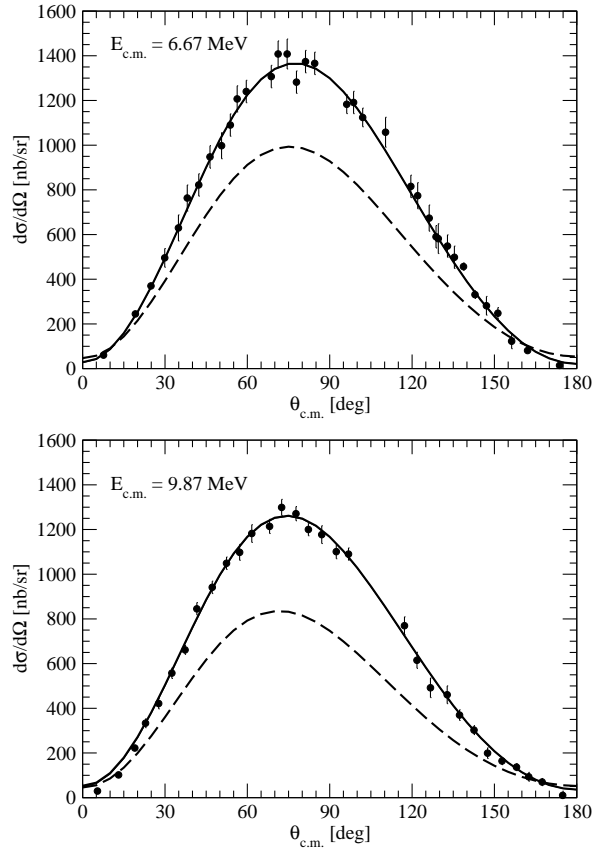}}
\caption{Differential cross section for 
$p-d$ capture at $E_{\rm c.m.}=6.67$ MeV (top panel) 
and $E_{\rm c.m.}=9.87$ MeV (lower panel). Notation as in 
Fig.~\protect\ref{fig:pd3MeV}. Thick and thin solid lines are 
on top of each other. 
The data are from Ref.~\protect\cite{Bel70}. } 
\label{fig:siu}
\end{figure}

\begin{figure}[t]
\centerline{
\epsfig{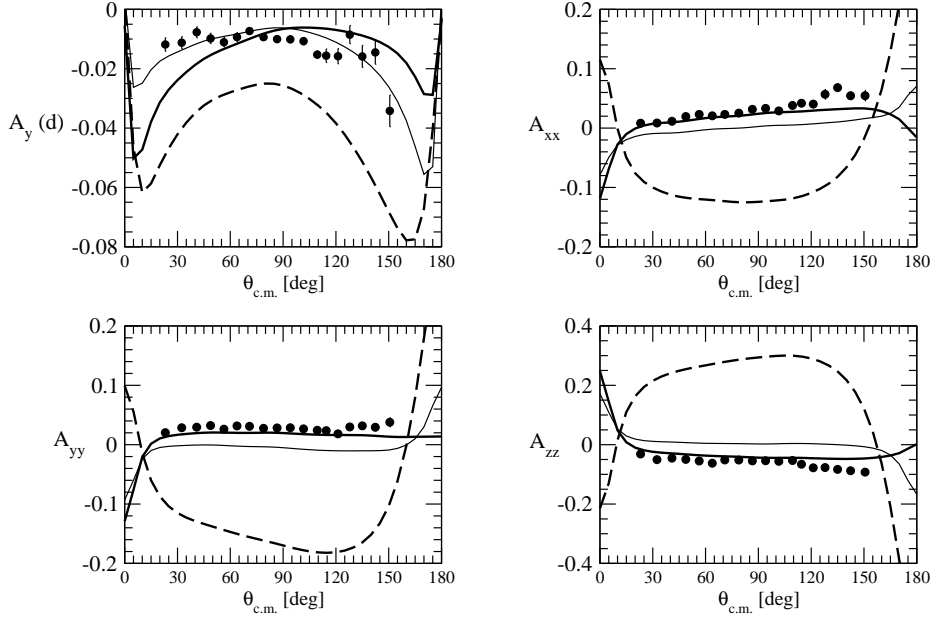}}
\caption{Deuteron tensor analyzing powers for  
$p-d$ capture at $E_{\rm c.m.}=5.83$ MeV. Notation as in 
Fig.~\protect\ref{fig:pd3MeV}. 
The data are from Ref.~\protect\cite{sagara}. } 
\label{fig:pd5MeV}
\end{figure}

The applications of our formalism for 
c.m. energies ranging from zero
to $2$ MeV, namely below the deuteron breakup threshold (DBT), were
already presented in Refs.~\cite{Viv96} and \cite{Viv99}.  Recently
we have extended the PHH technique, in order to compute also $p-d$ 
scattering wave
functions above the DBT~\cite{KRV01}.  We can therefore compute
$p-d$ capture observables at higher energies than previously
published. We present here in Fig.~\ref{fig:pd3MeV} 
a preliminary study at $E_{\rm c.m.}=3.33$ MeV,
where high-quality data, including differential cross sections,
vector and tensor analyzing powers~\cite{Knu92} exist. Results for 
differential cross sections at $E_{\rm c.m.}=6.67$ and 
9.87 MeV and deuteron tensor analyzing powers at 
$E_{\rm c.m.}=5.83$ MeV are given in Figs.~\ref{fig:siu} 
and~\ref{fig:pd5MeV}
respectively. The corresponding 
experimental data are taken from Refs.~\cite{Bel70} 
and~\cite{sagara}.

All the theoretical calculations 
reported  in this section have been obtained using  
the Argonne $v_{18}$ (AV18)~\cite{Wir95} two-nucleon and Urbana IX
(UIX)~\cite{Pud95} three-nucleon interactions. 
By inspection of Fig.~\ref{fig:pd3MeV}, 
we can conclude that our results (thin solid
lines) for the differential cross section and the observables 
$A_y$  and ${\rm i}T_{11}$ are in good agreement with the experimental data.
On the contrary, for the observables $T_{20}$ and
$T_{21}$, significant discrepancies can be observed. 
A similar situation was found for the $E_{\rm c.m.}=2$ MeV 
observables~\cite{Viv99}. The same situation is present also at higher 
energies. In fact the differential cross sections in Fig.~\ref{fig:siu} 
are well reproduced by our full calculation, as well as the $A_y(d)$ 
observable in Fig.~\ref{fig:pd5MeV}, 
which is basically proportional to ${\rm i}T_{11}$. Also 
at $E_{\rm c.m.}=5.83$ MeV, as can be seen from Fig.~\ref{fig:pd5MeV}, 
discrepancies 
are present for the deuteron tensor observables $A_{xx}$, $A_{yy}$ 
and $A_{zz}$, which are a linear combination of $T_{20}$, 
$T_{21}$ and $T_{22}$ (the latter observable is however rather small). 

The analysis of the $E_{\rm c.m.}=2$ MeV results in Ref.~\cite{Viv99} 
traced back the problem to an overprediction of the spin-flip electric
dipole $E_1$ reduced matrix elements (RME's) (namely, those arising from the
transitions where the $p-d$ spins in the incident channel are coupled
to $S=3/2$). In fact, 
when the same RME's are computed in the long wavelength
approximation (LWA) at leading order (thick solid line), 
the observables $T_{20}$ and $T_{21}$ are
better reproduced (see Figs.~\ref{fig:pd3MeV} and~\ref{fig:pd5MeV}). 
Interestingly, an analysis of the next-to-leading order terms in the LWA
performed in Ref.~\cite{Viv99}, has shown that they give a
sizeable contribution to the spin-flip $E_1$ RME's, showing the inadequacy
of the use of the leading order only for the calculation
of this small spin-flip transition matrix elements. The origin of the
discrepancies observed in the deuteron tensor polarization observables is
currently under investigation. 

\section{Muon Capture}
\label{sec:mu}
The $\mu^-$ weak capture on $^3$He can occur through three 
different hadronic channels:
\begin{eqnarray}
  \mu^- + ^3\!{\rm He} &\rightarrow& ^3{\rm H} +\nu_\mu ~~~(70\%)\ , 
  \label{eq:mucap} \\
  \mu^- + ^3\!{\rm He} &\rightarrow& n\,+\,d\,+\nu_\mu ~~~(20\%)\ , 
  \label{eq:mcnd} \\
  \mu^- + ^3\!{\rm He} &\rightarrow& n\,+\,n\,+p\,+\nu_\mu ~~~(10\%)\ . 
  \label{eq:mcnnp} 
\end{eqnarray}
The focus of the present section is on the first process~\cite{Mar02}.  
Some of
the nuclear physics issues in muon capture have been reviewed
recently in Ref.~\cite{Mea01}. When the triton polarization is not detected,
the differential capture rate for the reaction~(\ref{eq:mucap}) is given by
\begin{equation}
  {{\rm d}\Gamma \over {\rm d}(\cos\theta)} = {1 \over 2}\Gamma_0
  \Bigg[ 1+A_v P_v \cos\theta  
     + A_t P_t \left({3 \over 2}\cos^2\theta  -{1 \over 2}\right)
     + A_\Delta P_\Delta  \Bigg] \ , \label{eq:dgd} 
\end{equation}
where  $\Gamma_0$ is the total capture rate, $A_v$, $A_t$ and
$A_\Delta$ angular correlation parameters and $\theta$ is the angle
between the muon polarization and the leptonic momentum transfer.
The coefficients $P_v$, $P_t$ and $P_\Delta$ are linear combinations
of the probabilities $P(f,\,f_z)$  of
finding the $\mu^-\,^3$He system
in the total-spin state $|f\,f_z\rangle$,  and are defined
as~\cite{Con92,Hwa78}
\begin{eqnarray}
  P_v      &=&P(1,1)-P(1,-1) \ , \nonumber \\
  P_t      &=&P(1,1)+P(1,-1)-2\,P(1,0) \ ,  \label{eq:pol} \\
  P_\Delta &=&P(1,1)+P(1,-1)+P(1,0)-3\,P(0,0) \ .\nonumber
\end{eqnarray}
Therefore, $P_v$ and $P_t$ are proportional to the vector and tensor
polarizations of the $f$=1 state, respectively, while $P_{\Delta}$
indicates the deviation of the $f$=0 population density from its statistical
factor 1/4.  Because of the small energy splitting between the $f$=0 and $f$=1
hyperfine states (1.5 eV) compared to the $\mu^-\,^3$He binding energy,
and hence small deviation of $P(f,\,f_z)$ from its statistical value,
direct measurements of the angular correlation parameters are 
rather difficult~\cite{Mea01,Con92,Sou98}.

The results reported here have been obtained using either the AV18 and 
AV18/UIX interactions or the older Argonne $v_{14}$ (AV14)
two-nucleon interaction~\cite{Wir84} in conjunction with the the
Tucson-Melbourne (TM) three-nucleon interaction~\cite{Coo79}.
Note that both three-nucleon interactions have been adjusted 
to reproduce the triton binding energy.

Particular care has been put in the calculation of the $1s$ wave function
$\psi_{1s}(x)$ of the muon in the electric 
field of the $\het$ nucleus. Since the Bohr radius 
of the muonic atom is about 130 fm, i.e. much larger than the 
nuclear radius, it is well justified to approximate 
$\psi_{1s}(x)$ in the computation of the weak transition 
matrix elements with an average value $\psi_{1s}^{\rm av}$, defined as:
\begin{equation}
\psi_{1s}^{\rm av} = \frac{
\int\,d{\bf x}\,\, {\rm e}^{{\rm i}{\bf q}\cdot{\bf x}} \psi_{1s}(x)\,\rho(x)}
{\int\,d{\bf x}\,\, {\rm e}^{{\rm i}{\bf q}\cdot{\bf x}} \rho(x)} \ ,
\label{eq:psiav}
\end{equation}
where $\rho(x)$ is the $\het$ nucleus charge distribution
and $\bf{q}$ is the leptonic momentum transfer. Finally, it is 
common to introduce the factor ${\cal{R}}$ defined as~\cite{Wal75,Wal95}:
\begin{equation}
|\psi_{1s}^{\rm av}|^2 \equiv\,  {\cal {R}} |\psi_{1s}(0)|^2\,=\,
{\cal {R}}\,{(2\,\alpha\, m_{r})^3\over \pi} \ ,
\label{eq:psimu}
\end{equation}
where $\psi_{1s}(0)$ denotes the Bohr wave function
evaluated at the origin for a point charge $2 e$, 
and $m_r$ is the reduced mass of the $\mu^-\,^3$He system. 
The factor ${\cal {R}}$ 
has been calculated with the following procedure: for a 
given Hamiltonian model we have calculated $\rho(x)$. Knowing 
$\rho(x)$, the Poisson 
equation has been solved to obtain the static electric potential. 
Finally we have solved the Dirac equation and found 
$\psi_{1s}(x)$. Using Eqs.~(\ref{eq:psiav}) 
and~(\ref{eq:psimu}), we have calculated ${\cal {R}}$ for the three 
nuclear Hamiltonian models considered here. The results  
are given in Table~\ref{tb:r}. 
They are in rather good agreement with the values given in 
literature~\cite{Con92}. 

\begin{table}[t]
\caption{\label{tb:r} 
Values for ${\cal{R}}$ as defined in 
Eq.~(\ref{eq:psimu}) for three different Hamiltonian models $H$.}
\centerline{
\begin{tabular}{cc}
\hline
$H$       &  ${\cal{R}}$  \\ 
\hline
AV18      & 0.979 \\
AV18/UIX  & 0.980 \\
AV14/TM   & 0.980 \\
\hline
\end{tabular}
}
\end{table}
\begin{table}[b]
\caption{\label{tb:gapm} 
Capture rate $\Gamma_0$ (sec$^{-1}$) and angular
correlation parameters $A_v$, $A_t$, and $A_\Delta$
calculated using PHH wave functions corresponding
to the AV18, AV18/UIX and AV14/TM Hamiltonian
models are compared with the experimental results.  The theoretical
uncertainties, shown in 
parentheses, reflect the uncertainty in the
determination of the $N$$\Delta$ transition
axial coupling constant $g_A^*$. The experimental
values of $\Gamma_0$ and $A_v$ have been taken from Ref.~\cite{Ack98} 
and~\cite{Sou98}, respectively.
Here, we have assumed $R_{PS}=1$ in Eq.~(\ref{eq:gps2}).}
\centerline{
\begin{tabular}{crrrc}
\hline
Observable &    AV18       &  AV18/UIX    &  AV14/TM    & Expt.  \\ 
\hline
$\Gamma_0$ &   1441(7)     &   1484(8)    &   1486(8)   & 1496(4) \\
$A_v$      &   0.5341(14)  &   0.5350(14) &   0.5336(14)& 0.63(15) \\
$A_t$      & --0.3642(9)   & --0.3650(9)  & --0.3659(9) & \\ 
$A_\Delta$ & --0.1017(16)  & --0.1000(16) & --0.1005(17)& \\ 
\hline
\end{tabular}
}
\end{table}

Results for the capture rate $\Gamma_0$ and
angular correlation parameters $A_v$, $A_t$, and
$A_{\Delta}$ are presented in Table~\ref{tb:gapm}.  The uncertainty
(in parentheses) in the predicted values is due to
the uncertainty in the determination of  
the $N$$\Delta$ transition parameter $g_A^*$, as
discussed in the Introduction. The latter reflects the experimental
error in the Gamow-Teller matrix element of tritium
$\beta$-decay.

Inspection of Table~\ref{tb:gapm} shows that the
theoretical determination of the total capture rate $\Gamma_0$,
when the AV18/UIX and AV14/TM Hamiltonian models are used, 
is within 1 \% of the recent experimental result~\cite{Ack98}.
Furthermore,
the model dependence in the calculated observables
is very weak: the AV18/UIX and AV14/TM results
differ by less than 0.5 \%.  The agreement between
theory and experiment and the weak model dependence
mentioned above reflect, to a large extent, the fact that
both the AV18/UIX and AV14/TM Hamiltonian models
reproduce: i) the experimental binding energies as well
as the charge and magnetic radii~\cite{Mar98} of the
three-nucleon systems; ii) the Gamow-Teller matrix element in
tritium $\beta$-decay.  
In this respect, it is interesting to note 
that the capture rate predicted by the AV18 
Hamiltonian model is about 4 \% smaller
than the experimental value. The same result has been found with the 
two-nucleon AV14 Hamiltonian model~\cite{Mar02}.

The value for the angular correlation
parameter $A_v$ listed in Table~\ref{tb:gapm}
is also in reasonable agreement with the corresponding experimental
result which has however a rather large error. 
Note that the polarization 
observables are not sensitive to the inclusion of the three-nucleon 
force, as can be seen comparing the AV18 and AV18/UIX lines of 
Table~\ref{tb:gapm}.

\begin{table}[t]
\caption{\label{tb:gac} 
Effects of the inclusion of the two-body currents for the muon
capture rate $\Gamma_0$ (in sec$^{-1}$)  and angular correlation
parameters $A_v$, $A_t$, and $A_\Delta$.  The PHH wave functions are
obtained using the AV18/UIX Hamiltonian model.  The column  
labeled \lq\lq One-body\rq\rq~lists the contributions
associated with the one-body vector and axial charge
and current operators. The column labeled 
\lq\lq Mesonic\rq\rq~lists the results obtained
by including, in addition, the contributions from
meson-exchange mechanisms.  Finally the column labeled
\lq\lq$\Delta$\rq\rq~lists the results obtained by including
also the $\Delta$-excitation contributions.
 The experimental
vaues of $\Gamma_0$ and $A_v$ are taken from Ref.~\cite{Ack98}
and~\cite{Sou98}, respectively.
Here, we have assumed $R_{PS}=1$ in Eq.~(\ref{eq:gps2}).}
\centerline{	
\begin{tabular}{ccccc}
\hline
Observable &  One-body & Mesonic &  $\Delta$   & Expt. \\
\hline
$\Gamma_0$ &    1316   &  1384   &   1484   & 1496(4) \\
$A_v$      &   0.5749  &  0.5511 &   0.5350 & 0.63(15)\\
$A_t$      & --0.3565  &--0.3679 & --0.3650 & \\ 
$A_\Delta$ & --0.0686  &--0.0810 & --0.1000 & \\ 
\hline
\end{tabular}
}
\end{table}

The contributions of the different components of
the weak current and charge operators to the
observables are reported for the AV18/UIX model
in Table~\ref{tb:gac}. The column labeled 
\lq\lq One-body\rq\rq~lists the contributions associated with the one-body
terms of the vector and axial charge and current
operators, including relativistic corrections
proportional to $1/m^2$. 
The column labeled \lq\lq Mesonic\rq\rq~lists the results 
obtained by including, in addition, the 
contributions from two-body
vector and axial charge and current
operators, associated with pion-
and vector-meson-exchanges.  Finally, the column labeled \lq\lq
$\Delta$\rq\rq~lists the values of the observables obtained by
including the contributions arising from $\Delta$ excitation.

Among the observables, $\Gamma_0$ and $A_\Delta$
are the most sensitive to two-body contributions
in the weak current.  These are in fact crucial
for reproducing the experimental capture rate,
see Table~\ref{tb:gac}.

\begin{figure}[t]
\centerline{
\epsfig{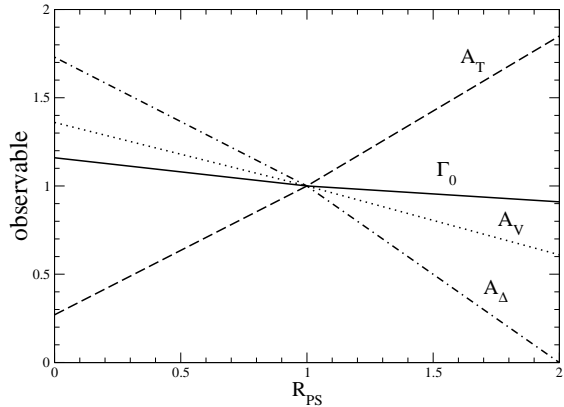}}
\caption{Variation of the capture rate $\Gamma_0$ and
angular correlation parameters $A_v$, $A_t$, and $A_\Delta$
with the parameter $R_{PS}$ entering the expression for the induced
pseudo-scalar coupling $G_{PS}$ given in Eq.~(\ref{eq:gps2}). The AV18/UIX 
PHH wave functions are used.  For each observable, the ratio 
between the result obtained with the given value of $R_{PS}$
and that one with the PCAC value $R_{PS}=1$
is plotted as function of $R_{PS}$.}
\label{fig:gps}
\end{figure}

An important motivation of the present work is to 
test the sensitivity of the muon capture
observables to the induced pseudo-scalar form factor
$G_{PS}$ and, eventually, infer its value from the $\Gamma_0$ measurement.
Therefore,  we have repeated the calculation 
using AV18/UIX PHH wave functions and several different 
values of the parameter $R_{PS}$, defined in Eq.~(\ref{eq:gps2}).
The variation of each observable in terms of $R_{PS}$
is displayed in Fig.~\ref{fig:gps}.
The angular correlation parameters, in particular $A_t$ and
$A_\Delta$, are more sensitive to changes in $R_{PS}$ than
the total capture rate, as first pointed out in Ref.~\cite{Hwa78}.
A precise measurement of these polarization observables
could therefore be useful to ascertain the extent to which
the induced pseudo-scalar form factors deviate from their PCAC values.

By enforcing perfect agreement between the
experimental and  theoretical values, taken with their uncertainties,
for the total capture rate $\Gamma_0$, it is possible to obtain an
estimate for the range of values allowed for $R_{PS}$, and we have
found 
\begin{equation}
R_{PS}=0.94 \pm 0.06 \ .
\label{eq:gpsth}
\end{equation} 
This 6 \% uncertainty is significantly smaller than that found
in previous studies~\cite{Con92,Con95,Gov00}.  This
substantial reduction in uncertainty can be traced back
to the procedure used to constrain the (model-dependent)
two-body axial currents discussed in the Introduction.

\section{Summary and Outlook}
\label{sec:setc}

We have reported new calculations of
$p-d$ radiative capture observables at energies
above the DBT, and
of observables in the process $^3$He($\mu^-$, $\nu_\mu$)$^3$H.
These calculations have been based on the Argonne
$v_{18}$ two-nucleon and
Urbana IX three-nucleon
interactions. For the muon capture reaction also the older Argonne 
$v_{14}$ two-nucleon and Tucson-Melbourne three-nucleon interactions 
have been used. Accurate bound and
continuum wave functions have been  obtained with the
PHH method.  The model for the EM and weak transition
operators has been taken to consist of one- and
two-body components.  In recent studies, this theory has been shown
to correctly predict the static properties
of the three-nucleon systems~\cite{Mar98}, as well as
their associated elastic and transition
electromagnetic form factors. 

A satisfactory description of measured $p-d$ capture observables at
$E_{\rm c.m.}=3.33$ MeV has emerged with the exception of the $T_{20}$ and
$T_{21}$ tensor analyzing powers.  Interestingly, the very same
discrepancies were observed also below the DBT, namely at $E_{\rm c.m.}=2$
MeV. Similarly, at higher values of $E_{\rm c.m.}$, differential 
cross sections and $A_y(d)$ seem to be well reproduced, while 
the tensor analyzing powers $A_{xx}$, $A_{yy}$ and $A_{zz}$ at 
$E_{\rm c.m.}=5.83$ MeV present the same discrepancies mentioned above.

In order to clarify the origin of these discrepancies, we plan
to extend the calculation of $p-d$ capture observables at even 
higher energies, and to investigate alternative models for
short-range two-body EM currents.

In regard to the muon capture process, we have found that
the predicted total rate is in agreement
with the experimental value, and has only
a weak model dependence: the AV18/UIX and AV14/TM results differ
by less than 0.5 \%.  As discussed above, the weak model dependence 
comes about because both Hamiltonians reproduce the binding
energies, charge and magnetic radii of the three-nucleon systems, and 
the Gamow-Teller matrix element in tritium $\beta$-decay.

It is important to note that, if the contributions
associated with two-body terms in the axial
current were to be neglected, the predicted capture
rate would be 1316 (1318) sec$^{-1}$ with AV18/UIX (AV14/TM), 
and so two-body mechanisms
are crucial for reproducing the experimental value.
The present work shows that the
procedure adopted for constraining these two-body
contributions leads to a consistent description of
available experimental data on weak transitions in
the three-body systems.  It also corroborates the
robustness of our recent predictions for the cross
sections of the proton weak captures on $^1$H~\cite{Sch98}
and $^3$He~\cite{Mar01}, which were obtained
with the same model for the nuclear weak current.

Finally, it would be interesting to extend our investigation to the 
$^3$He($\mu^-$,$\nu_\mu$)$n$$d$ and $^3$He($\mu^-$,$\nu_\mu$)$n$$n$$p$
processes, both of which have been studied
experimentally in Ref.~\cite{Kuh94} and
theoretically in Ref.~\cite{Ski99}.  
Work along these lines is vigorously being pursued.

The work of R.S. is supported by the U.S. Department of Energy
contract number DE-AC05-84ER40150 under which the Southeastern
Universities Research Association (SURA) operates the
Thomas Jefferson National Accelerator Facility.

\end{document}